\documentclass[prb,aps,twocolumn]{revtex4-1}
\usepackage{graphicx,amsmath,color,soul}
\usepackage{epsfig}

\usepackage{epstopdf}
\usepackage[export]{adjustbox}

\usepackage[colorlinks,urlcolor=blue,citecolor=blue,linkcolor=blue]{hyperref}
\begin{document}
\UseRawInputEncoding
\title{Sensing food quality by silicene nanosheets : a Density Functional Theory study}

\author{Madhumita Kundu}
\email{kmadhumita@iitg.ac.in}
\affiliation{Department of Physics, Indian Institute of Technology Guwahati, Assam 781039, India}

\author{Subhradip Ghosh}
\email{subhra@iitg.ac.in}
\affiliation{Department of Physics, Indian Institute of Technology Guwahati, Assam 781039, India}

\begin{abstract}
Volatile organic compounds (VOCs) emitted by food products are considered markers for assessing quality of food. In this work, first-principles Density Functional Theory (DFT) and Non-equilibrium Green's function (NEGF) methods have been employed to model chemo-resistive gas sensor based on two-dimensional silicene based nanosheets that can sense the six different VOCs emitted by standard food products. Our calculations with unpassivated and flourine passivated silicene(F-silicene) sheets as sensor materials show that flourine passivated silicene has significantly better sensitivity towards all six VOC molecules (Acetone, Dimethylsulfide, Ethanol, Methanol, Methylacetate and Toluene). Moreover, flourinated silicene sensor is found to be capable of separately recognising four VOCs, a much better performance than r-GO used in a recent experiment. We analyse the microscopic picture influencing sensing capabilities of un-passivated and fluorinated silicene from the perspectives of adsorption energy, charge transfer and changes in the electronic structure. We find that better sensing ability of fluorinated silicene nanosheet can be correlated with the changes in the electronic structures near the Fermi level upon adsorption of different VOCs. The results imply that passivated silicene can work better as a sensor than r-GO in case of generic food VOCs. The results are important since modelling of various two-dimensional nano-sensors can be done in the similar way for detection of more complex VOCs emitted by specific food products.
\end{abstract}

\maketitle

\section{Introduction}
\vspace{-0.3cm}
Over past three decades, nano-science and technology have been omnipresent in common man's life. This is due to revolutionary advances in nano-electronics \cite{nanoelec} and nano-medicines \cite{nanomed}, among other applications. Since food and it's consumption is an integral part of human life, the quality and safety of food products is one of the key indices to gauge improvements in standards of living. Quite naturally, nano-science based applications have found their way into global industry surrounding food and helped develop various parameters related to food quality and safety \cite{nanofood,nanofood1}. This has been possible, thanks to advancements in intelligent packaging \cite{pack1,pack2,pack3,pack4} that tracks the trajectory of food history and quality from producer to consumer, and takes preventive measures to reduce wastage \cite{pack5}. The technology is based on gas sensors that can detect and distinguish chemicals released during the stages of maturation. The quality of food product, ascertained by its level of deterioration is often assessed by assembling the gas sensors into arrays leading to e-nose, an electronic machine designed to detect and discriminate among complex odours \cite{enose}. The odours are primarily due to volatile organic compounds (VOCs) released by food in the open space of the package. The sensor array can analyse the nature and content of the VOCs and determine the stage of deterioration. VOCs, therefore, are considered as markers of food spoilage.

Gas sensors based on nanomaterials are the most promising candidates for this purpose as high surface to volume ratio, tunability of electronic structures upon adsorption of gas molecules, high sensitivity and selectivity, fast response time, low power consumption, small size of nanomaterials work as advantages. Graphene, the prototype two-dimensional (2D) material has been extensively explored for gas-sensing and bio-sensing applications \cite{gas,gas1,gas2,gas3,gas4,gas5}. Graphene based nanomaterials have also been used to detect and differentiate various VOCs for breath analysis where they act as bio-markers for several terminal diseases \cite{breath} and to remove pollutants like NO$_{2}$ and SO$_{2}$ from atmosphere \cite{pollutant}. However, the sensors based on pristine graphene have limited sensitivity and selectivity towards diverse VOCs \cite{breath}. One way to overcome such limitation is its functionalisation with organic compounds,molecules,conducting polymers and introduction of dopants or defect engineering \cite{graphenenanosensors,graphenenanosensors1,graphenenanosensors2,graphenenanosensors3,graphenenanosensors4}.

Due to the limitations posed by pure graphene and the structural complications associated with its various composites, researchers, particularly ones performing modelling and simulations, have explored other materials from 2D family as potential gas sensors applicable to a wide spectrum of VOCs. A plethora of first-principles Density Functional Theory \cite{dft} based simulations have been carried out in recent times with pristine compounds like MoS$_{2}$ \cite{tian2017engineering}, phosphorene \cite{kou2014,zhang2015first,sun2019blue}, black phosphorous \cite{ou2019superior}, HfTe$_{2}$ \cite{johari} and BC$_{6}$N \cite{vocbc6n} as sensors for VOC and hazardous gas detection.

Graphene based nanocomposites have recently started to be used in food industry for variety of purposes associated with food safety \cite{graphenefood}. With regard to sensing VOCs from food products, chemoresistive responses of reduced Graphene oxide (r-GO) and it's nanocomposites with polythiopene have been experimentally measured upon exposure to flows of VOCs\cite{meatexpt}. Methanol (CH$_{3}$OH), Ethanol (C$_{2}$H$_{5}$OH), Acetone ((CH$_{3}$)$_{2}$OC), Methyl acetate ((CH$_{3}$)$_{2}$OCO), Dimethylsulfide ((CH$_{3}$)$_{2}$S)  and Toluene (C$_{6}$H$_{5}$CH$_{3}$) vapours were tested since they are the prominent VOCs released from meat \cite{meatvoc}, fruits \cite{fruitvoc} and vegetables \cite{vegvoc}, making them markers of food spoilage. Sensitivity and selectivity of the sensor material were decided by measuring the resistance relative amplitude defined as $A_{r}=\frac{R-R_{0}}{R_{0}}$, where $R(R_{0})$ is the resistance of the sensor in presence of a particular VOC(nitrogen) vapour. It was found that while pristine r-GO is unable to differentiate between the VOCs, it's nanocomposite would. No further comprehensive study on food VOCs using 2D materials as sensors is available till date. However, several first-principles DFT calculations on interactions of VOCs emitted from different fruits with various 2D sensing materials like Germanene, Stanene and silicene have been carried out \cite{germanenebanana,stanenebanana,silicenepearfruit}. These studies, although provided some insights into the interplay of orbital hybridisation, charge transfer, work function and subsequent influence of it's response when interacted with different organic molecules, the crucial aspect of device integrability was not addressed. As an alternative to Graphene, silicene, among the single-component 2D materials, has shown substantial promises. In contrast to $sp^{2}$-hybridised Graphene, silicene has a mixture of $sp^{2}$ and $sp^{3}$ hybridisation and hence exhibits a buckled structure. The buckling turns out to be advantageous over planar structure of Graphene as it provides tunability in the band gap upon application of external electric field. Accordingly, silicene based Field Effect Transistors (FETs) have been explored. It turns out that they demonstrate better efficiency than Graphene based FETs \cite{ghosal}. The material has been extensively investigated with regard to bio-sensing applications, through computational modelling. To this end, silicene nanosheets and nanoribbons as bio-sensors have been investigated for DNA sequencing \cite{hussain,henry,alesheikh,li2020} and sensing of important bio-molecules like uric acid \cite{tarun}, paracetamol \cite{saikia} and pyrazinamide, the organic molecule used to treat tuberculosis \cite{saikia1}. silicene nanosheets have also been used for sensing of hazardous gases \cite{prasongkit2015highly}. The band structure of silicene changes significantly upon surface passivation by functional groups. Hydrogenation that is passivation of surface dangling bonds by hydrogen changes the zero gap semiconductor to an insulator \cite{hydrogenation} while fluorine functionalisation yields a moderate band gap of 0.45 eV \cite{jia2015,zaminpayma2016}. This poses a possibility of band gap modulation upon adsorption of different VOCs and its connection to differential responses, giving rise to discrimination between VOCs, the desired goal. This, certainly is an advantage over graphene or r-GO as sensing material. 

In what follows, in this work, using DFT and non-equilibrium Green's function (NEGF) \cite{negf}  technique, we have investigated the changes in the structural and electronic properties as well as sensing capabilities of pristine and Fluorine (-F) functionalised silicene (F-silicene) nanosheets upon adsorption of six prominent VOCs mentioned above that are indicators of degradation of food products. The modifications in the electronic structures, associated charge transfers and work functions of the gas adsorbed nanosheets as well as the differences in these properties/parameters between bare and functionalised nanosheets provide necessary insights into their comparative capabilities as sensors to ascertain the degree of deterioration of food. The ideas obtained are tested by simulating the sensing device and calculating transport parameters that are directly related to the experimental quantities like sensitivity and selectivity. We find that both sensitivity and selectivity of silicene increase upon -F functionalisation at a moderate bias voltage. This demonstrates that functionalised silicene can be an sensor, alternative to Graphene based ones, in addressing quality of food products.

%
\begin{figure}[t]
\begin{center}
\includegraphics[width=0.485\textwidth]{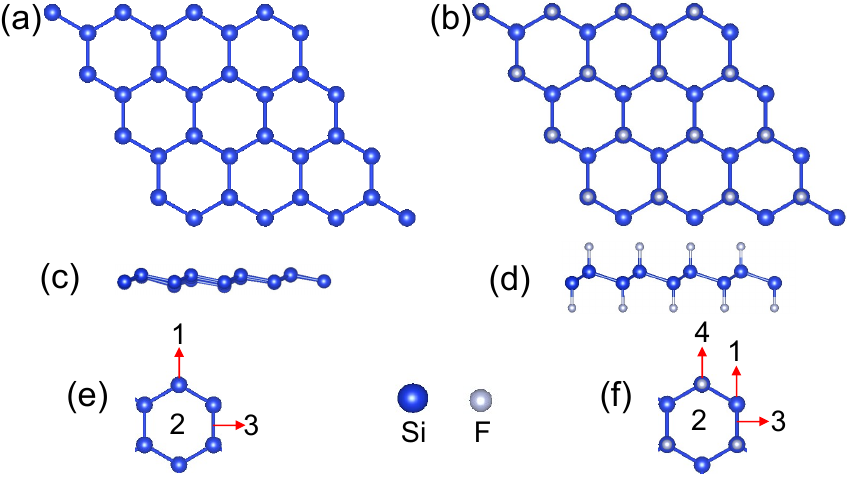}
\caption{Relaxed structure of silicene: (a)top view, (c)side view. Relaxed structure of  F-silicene: (b) top view and (d) side view. Sites of adsorption for (e) silicene and (f)F-silicene are shown.}  \label{Fig:1}
\end{center}
\end{figure}

\vspace{-0.2cm}
\section{COMPUTATIONAL DETAILS}
\vspace{-0.3cm}
Important physical quantities at equilibrium and the electronic structures associated with interactions due to adsorption of VOC molecules on silicene nanosheets are calculated using DFT with Projector Augmented Wave (PAW) method \cite{blochl1994projector} as implemented in Vienna Ab initio simulation package(VASP) \cite{kresse1996efficiency, feng2020predicting,kresse1999ultrasoft}.Exchange-correlation part of the Hamiltonian is approximated using the Generalized Gradient Approximation (GGA) with the Perdew-Burke-Ernzerhof (PBE) parametrisation \cite{perdew1996generalized}. Long-range van der Waals (vdW) interactions are considered by employing corrections proposed by Grimme\cite{grimme2006semiempirical}. $4 \times 4 \times 1$ fully relaxed supercells are used for the pristine and -F functionalised silicene nanosheets (Figure \ref{Fig:1}). In order to avoid interaction between two layers along $z$-direction, a vaccum of 30 \AA  is used. The geometry optimisation of the composite system of molecule adsorbed on the nanosheet is carried out using the conjugate gradient method. Plane wave basis set upto 550 eV is considered. Convergence criteria for total energy and force are kept at $10^{-6}$ eV and $0.04$ eV/\AA, respectively. A $4 \times— 4 \times— 1$ Monkhorst-Pack \cite{monkhorst1976special} $k$-point grid is used to sample the Brillouin zone for self-consistent calculations. A $17 \times 17 \times 1$ $k$-mesh is used for calculations of electronic structures.

The physisorption energy provides quantitative and qualitative information about the interaction strength between the adsorbent and the adsorbate. The adsorption energy of a single VOC molecule on the nanosheet can be expressed as:
\begin{align}
E_{\text{ad}}=\left[E_{\text{nanosheet+VOC}}-E_{\text{nanosheet}}-E_{\text{VOC}}\right]
\end{align}
where $E_{\text{nanosheet+VOC}}, E_{\text{nanosheet}}$ and $E_{\text{VOC}}$ are total energies of VOC-nanosheet complex, pristine nanosheet and the isolated VOC molecule, respectively. 

To assess the performances of the silicene based sensor devices, electronic transport through the VOC- nanosheet complexes are studied by the NEGF method in conjunction with DFT as implemented in the TRANSIESTA \cite{brandbyge2002density} package. The electrical current through the device with two metal electrodes under a finite bias voltage $V_{b}$  can be calculated using the Landauer-Buttiker formula as follows \cite{meir1992landauer, buttiker1988coherent, landauer1957spatial}:
\begin{align}
I(V_b)= \frac{2e}{h} \int T \left(E,V_b \right) \left(f_L \left(E-\mu_L \right)-f_R \left(E-\mu_R \right) \right)
\end{align}
$e$ represents the electronic charge, h is Planck's constant,$T(E, V_b)$ the electronic transmission function and $f_L(E - \mu_L)$ and
$f_R(E - \mu_R)$ are the Fermi functions in the left ($L$) and right ($R$) electrodes, respectively. $\mu_{L}$ and $\mu_{R}$ are the chemical potentials of the left and right electrodes, respectively. The bias energy window is between $\mu_{L}=E_{f}-eV_{b}/2$ and $\mu_{R}=E_{f}+eV_{b}/2$; $E_{f}$, the average Fermi energy, is given by $\left(\mu_{L}+\mu_{R} \right)/2$. The transmission coefficient $T(\epsilon)$ of electrons with incident energy $\epsilon$ is given by
\begin{align}
T(\epsilon) =  \Gamma_L \left(\epsilon \right)\mathcal{G} \left(\epsilon \right)\Gamma_R \left(\epsilon \right)\mathcal{G}^{\dagger} \left(\epsilon \right) 
\end{align}
Here, $\Gamma_R \left(\epsilon \right)$ and $\Gamma_L \left(\epsilon \right)$ are the coupling matrices for the right and left electrodes, respectively. $\mathcal{G}\left(\epsilon \right)$ represents the Green's function for the system. Nanosheets of size $5 \times 3 \times 1$ and $k$-point grids of size  $10 \times 1 \times 100$ are used for all transport calculations.

\vspace{-0.2cm}
\section{Results and Discussions} 
\subsection{Structural properties}
In Figure \ref{Fig:1} (a) and (c), the top and side views of the optimised pristine silicene sheet are shown. Corresponding structures for F-silicene are shown in \ref{Fig:1} (b) and (d). The calculated lattice constant (Si-Si bond length) of pristine silicene sheet is 3.84(2.27) \AA, in good agreement with existing results \cite{hussain1,zaminpayma2016}. In case of F-silicene, the optimised Si-Si bond length doesn't change much; the optimised Si-F bond length is 1.6 \AA, in agreement with other works \cite{zaminpayma2016,lusif}. The buckling parameter in either case is 0.47 \AA, in excellent agreement with the existing result of 0.4 \AA \cite{bucklingsilicene}.

\begin{figure*}
\includegraphics[width=0.9\textwidth]{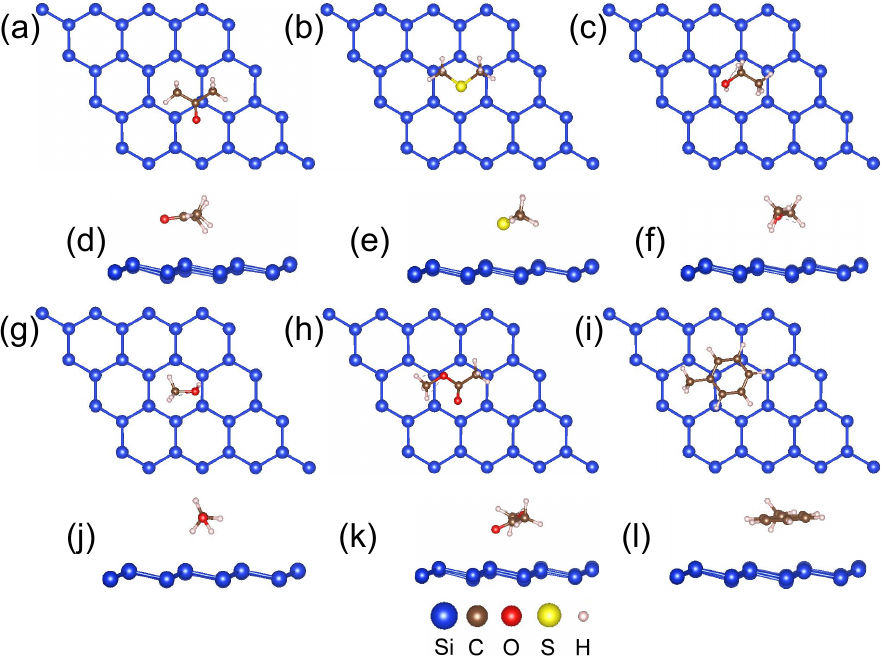}
\caption{(a-c) Top and (d-f) side views of optimised geometries after (a,d) Acetone, (b,e) Dimethylsulfide, and (c,f)Ethanol physisorbed on silicene nanosheet. (g-i)Top and (j-l) side views of optimised geometries after (g,j) Methanol, (h,k) Methylacetate and (i,l) Toluene physisorbed on silicene nanosheet.}  \label{Fig:2}
\end{figure*}

\begin{figure*}
\includegraphics[width=0.9\textwidth]{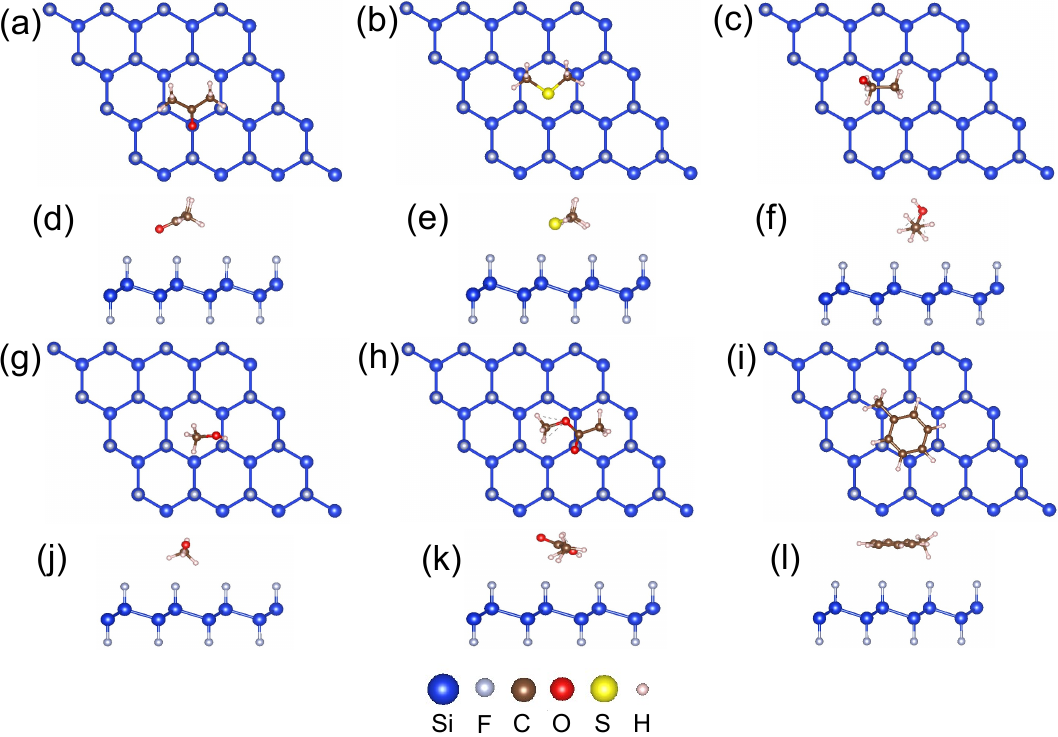}
\caption{(a-c) Top and (d-f) side views of optimised geometries after (a,d) Acetone, (b,e) Dimethylsulfide, and (c,f)Ethanol physisorbed on F-silicene nanosheet. (g-i)Top and (j-l) side views of optimised geometries after (g,j) Methanol, (h,k) Methylacetate and (i,l) Toluene physisorbed on F-silicene nanosheet.}  \label{Fig:3}
\end{figure*}

There are three sites of adsorption, marked as 1-3 in Figure \ref{Fig:1} (e) in pristine silicene sheet. For F-silicene, there is one extra site available due to -F functional site (Figure \ref{Fig:1}(f)). For each of the six VOC molecule, we optimise the geometry of the nanosheet-VOC complex and obtain the preferred site of binding and the orientation of the gas molecule with respect to the nanosheet. The sites of adsorption and the orientations of the six VOC molecules with respect to pristine silicene(F-silicene) sheet are shown in Figure \ref{Fig:2}(Figure\ref{Fig:3}). Associated binding energies, sites of adsorption as numbered in Figure \ref{Fig:1} and the distances of the molecules from the sheets are tabulated in Table \ref{TABLE1}. 

In order to get insights into the efficiency of sensing, the first thing to study is the adsorbent-adsorbate interaction of a potential gas sensor. The common wisdom is that the interaction should be strong enough to hold the gas molecule on the sensor. At the same time, the interaction strength should be such that the molecule should be easily removable without affecting other properties of the sensor. The strength of interaction can be gauged from $E_{\text{ad}}$. In our case, we find the following trends: first, the interaction strength for all six VOCs is weaker with F-silicene as the sensor material and second, irrespective of the 2D sensor material, the interaction strength in the ascending order is Methanol, Ethanol, Acetone, Methylacetate, Dimethylsulfide, Toluene. The adsorption energy varies over a wide range: from -0.17 eV for Methanol on F-silicene to -0.64 eV for Toluene on silicene. The values suggest that except Toluene, all other molecules are moderately physisorbed on the silicene and F-silicene nanosheets. The adsorption energies compare well with existing results obtained for different 2D adsorbent materials. Our results on Acetone are either better in case of adsorbent being MoS$_{2}$(-0.14 eV)\cite{tian2017engineering}, blue phosphorene (-0.25 eV)\cite{sun2019blue} or agree closely when the adsorbent is black phosphorous(-0.32 eV)\cite{ou2019superior} and BC$_{6}$N (-0.3 eV)\cite{vocbc6n}. For Ethanol, our results agree closely too (-0.21 eV for MoS$_{2}$ \cite{tian2017engineering}, -0.2 eV for blue phosphorene \cite{sun2019blue}, -0.24 eV for black phosphorous \cite{ou2019superior} and -0.38 eV for BC$_{6}$N \cite{vocbc6n}). In case of Toluene and Methanol, our results differ substantially when compared with results with BC$_{6}$N adsorbent (-0.91 eV for Toluene and -0.31 eV for Methanol) \cite{vocbc6n}. However, with black phosphorous as the nanosheet, our results are in good agreement for Toluene (-0.5 eV) \cite{ou2019superior}. It is, therefore, expected that the performance of silicene and F-silicene as sensors will be comparable to the standard 2D gas sensors. 

In order to understand the trends in the adsorption energies, we look at the optimised structures of the VOC molecules on two different nanosheets. From Table \ref{TABLE1}, we find that except Methanol and Methylacetate, the molecules prefer the same adsorption site, irrespective of the adsorbent. Acetone prefers one of the Si-sites of the hexagon. Dimethylsulfide  prefers the centre of the hexagon while Toluene prefers a bridge site. Ethanol, Methanol and Methylacetate prefer the centre of the hexagon(bridge site) when the sensing material is pristine silicene (F-silicene). When physisorbed on pristine silicene, the acetone molecule prefers an orientation almost parallel to the nanosheet, with the oxygen atom closest to the sheet and at a distance of 3.43 \AA. When the adsorbent is F-silicene, the molecule prefers the same orientation but is at a greater distance of 3.97 \AA  from the nanosheet. That the interaction distance $D$ increases substantially when the nanosheet is F-silicene, is also observed in case of Methylacetate and Toluene. For the other three, the trend is opposite. While Toluene molecules are adsorbed in an orientation parallel to the sheet for both nanomaterials with H atom closest to the adsorbent, Methylacetate is adsorbed with a tilted orientation with respect to the nanosheets. When the nanomaterial is pristine silicene, the O atom is the closest to it. The orientation changes in case of F-silicene, bringing H closest to the adsorbent. The other three molecules prefer tilted orientations with respect to both adsorbents. While H atom remains the closest to the nanosheet in case of Ethanol and Methanol, it is the S atom in case of Dimethylsulfide. 

Among the six VOCs, Toluene has the maximum number of atoms and is devoid of O. Dimethylsulfide is the only other molecule having no O atom. This explains highest and the next highest $E_{\text{ad}}$ for these two. The hierarchy with respect to $E_{\text{ad}}$ among the four VOCs containing O can be explained in terms of number of atoms each molecule has. $E_{\text{ad}}$ is higher in case of the molecule with greater number of atoms. The hierarchy among these four can also be understood based upon the orientation of the molecules with respect to the nano-sheets and the $D$ value. O atoms are closest to the adsorbent in case of Acetone and Methylacetate while H atoms are the closest for Ethanol and Methanol. Due to weaker Si-H bonds in comparison to Si-O bonds, in case of pristine silicene the sensing material, $E_{\text{ad}}$ for Acetone and Methylacetate adsorption on pristine silicene are larger. The larger $D$ in case of Acetone explains smaller $E_{\text{ad}}$ for Acetone than that of Methylacetate. Larger $E_{\text{ad}}$ of Ethanol, in comparison to Methanol, can be explained the same way. 

The decrease of $E_{\text{ad}}$ when silicene nanosheet is functionalised with -F cannot be explained in terms of $D$. This is because an inverse relationship between $D$ and $E_{\text{ad}}$, as expected, is observed only for Acetone, Methylacetate and Toluene adsorption on F-silicene sheet. This can be understood the following way: when silicene surface is functionalised with -F, very strong Si-F covalent bonds are formed. Since -F is highly electronegative, a protective layer is created on the silicene surface that prevents silicene to react with VOC molecules. The strong Si-F bonds too make it difficult for VOC molecules to react. Thus the binding of VOCs is relatively weak when the nanosheet is -F passivated silicene.

\begin{table*}
\caption{\label{TABLE1} Site of adsorption, Adsorption Energy ($E_{\text{ad}}$) and equilibrium distance between 2D sheet and VOC Molecule($D$) for the systems considered in this work.}
\begin{adjustbox}{width=0.5\textwidth}
\small
\begin{tabular}{c|cccccccc}
 \hline
 Adsorped & 2D & &Adsorption  &  &\boldmath{$E_{\text{ad}}$ }& & &\boldmath{$D$ } \\
 molecule & sheet & & site & &(eV) & & & (\AA)\\
 \hline
  & silicene  & &1 & & -0.34 & & & 3.43 \\
 Acetone ((CH$_{3}$)$_{2}$CO) & & & & & & & & \\
 \cline{2-9}\\
 & F-silicene & & 1 & & -0.32 & & & 3.97 \\
 \hline
 & silicene & & 2 & & -0.42 & & & 3.55 \\
 Dimethylsulfide ((CH$_{3}$)$_{2}$S) & & & & & & & & \\
  \cline{2-9}\\
 & F-silicene & & 2 & & -0.35 & & & 3.09 \\
 \hline
 & silicene & & 2 & & -0.32 & & & 2.76  \\
 Ethanol (C$_{2}$H$_{5}$OH) & & & & & & & &\\
  \cline{2-9}\\
 & F-silicene & & 3 & & -0.24 & & & 2.6 \\
 \hline
 & silicene & & 2 & & -0.25 & & & 2.89 \\
 Methanol (CH$_{3}$OH)& & & & & & & &\\
  \cline{2-9}\\
 & F-silicene & & 3 & & -0.17 & & & 2.3 \\
 \hline
 & silicene & & 2 & & -0.39 & & & 2.92 \\
 Methylacetate ((CH$_{3}$)$_{2}$OCO) & & & & & & & &\\
  \cline{2-9}\\
 & F-silicene & & 3 & & -0.35 & & & 3.36 \\
  \hline\\
 & silicene & & 3 & & -0.64 & & & 3.18 \\
 Toluene (C$_{6}$H$_{5}$CH$_{3}$) & & & & & & & &\\
  \cline{2-9}\\
 &F-silicene & & 3 & & -0.48 & & & 3.91 \\
 \hline
\end{tabular}
\end{adjustbox}
\end{table*}
\begin{figure*}[t]
\begin{center}
\includegraphics[width=1.0\textwidth]{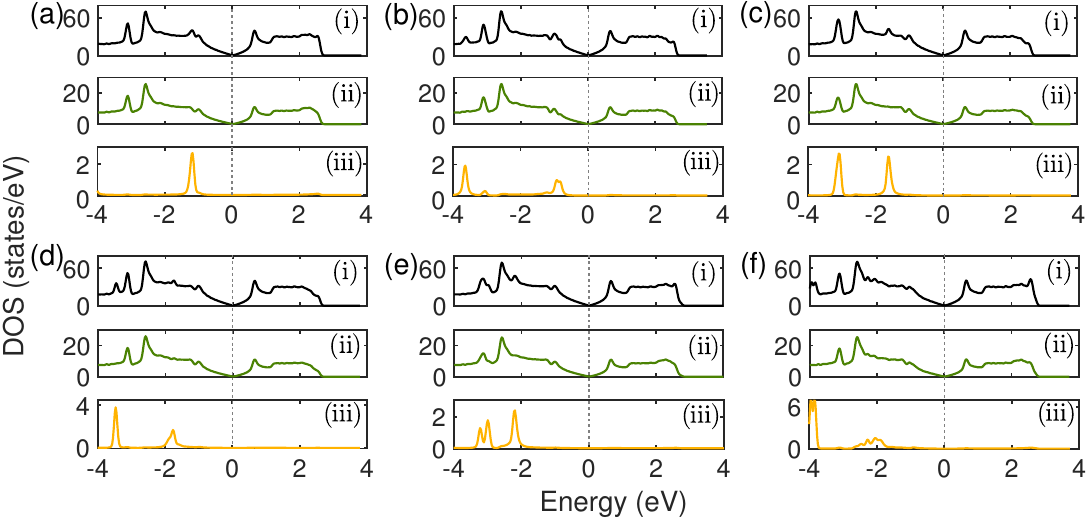}
\caption{i) Total DOS (ii) DOS of silicene and (iii) DOS of VOC molecule in (a)Acetone-silicene (b)Dimethylsulfide-silicene (c)Ethanol-silicene (d)Methanol-silicene (e)Methylacetate-silicene and (f)Toluene-silicene systems. } \label{Fig:4}
\end{center}
\end{figure*}
\begin{figure*}[t]
\begin{center}
\includegraphics[width=1.0\textwidth]{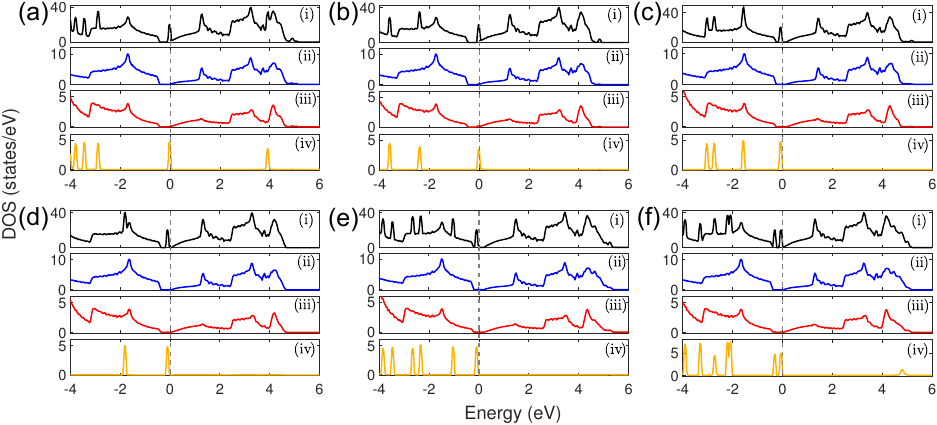}
\caption{(i) Total DOS (ii) DOS of silicene (iii) DOS of F and (iv) DOS of VOC molecule in (a)Acetone-F-silicene (b)Dimethylsulfide-F-silicene (c)Ethanol-F-silicene (d)Methanol-F-silicene (e)Methylacetate-F-silicene and (f)Toluene-F-silicene systems.}  \label{Fig:5}
\end{center}
\end{figure*}

\subsection{Electronic Structure}
More profound understanding of VOC adsorption on the nanosheets and their possible consequences with regard to sensing abilities of these 2D materials can be obtained by analysing the electronic structures of the VOC-nanomaterial complexes. In Figures \ref{Fig:4} and \ref{Fig:5} we show the total and component-projected densities of states (DOS) of the systems considered in this work. The Total densities of states of the VOC complexes with silicene as the adsorbent (Figure \ref{Fig:4}) show that there is very little change in the electronic structure, particularly close to the Fermi level, upon adsorption of different VOCs. The major difference in the total DOS occurs around -1 eV to -2 eV and between -3 eV to -4 eV, due to the states of the VOC molecules. For all cases, the VOC states are contributed by the valence states of all constituent atoms. The structure around -1 eV for Acetone-silicene complex (Figure \ref{Fig:4}(a)) is due to the hybridisation of Si, C and O $p$ and H $s$ states. The presence of two -CH$_{3}$ functional group symmetrically networked with the C-O bond leads to near equal contributions of all atoms of the molecule. In case of adsorbed Dimethylsulfide, its states near -1 eV are contributed largely by the $3p$ states of S and $s$ states of the two H atom connected to the two C atoms and are closer to the surface of the nanosheet. The deeper states around -4 eV have contributions from valence orbitals of all atoms in the molecule. For Ethanol, upon adsorbed in silicene, the states around -2 eV are contributed by $p$ orbitals of  O, $p$ orbitals of C and the $s$ orbitals of two H attached to them, situated closer to the sheet after relaxation (Figure \ref{Fig:2}(c) and (f)). For the adsorbed Methanol, the states around -2 eV are similarly contributed by the O $p$ and $s$ states from the two H atoms attached to the C atom (Figure \ref{Fig:2}(g) and (j)) while $p$ states of O closer to silicene surface and those of two C atoms connected (Figure \ref{Fig:2}(h) and (k)) are the major contributors to the states near -2 eV for Methylacetate-silicene complex. The DOS of toluene adsorbed silicene is somewhat different from the rest. This is due to weak contributions from Toluene around -2 eV. The major contributions from the constituents of Toluene molecule are located deeper in the valence band. The features in the densities of states imply that the charge transfer between the VOC and the nanosheet will be maximum for Acetone and least for Toluene VOC. Also, since there is no noteworthy changes in the electronic structures upon adsorption of different VOCs, the selectivity of silicene sensor may be poor. 

\begin{figure*}[t]
\begin{center}
\includegraphics[width=1.0\textwidth]{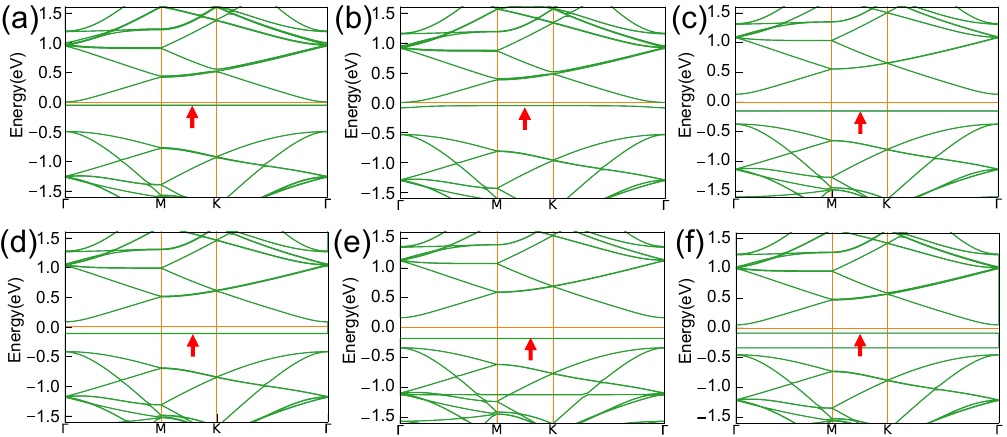}
\caption {Band Structure of (a) Acetone-F-silicene (b)Dimethylsulfide-F-silicene (c)Ethanol-F-silicene (d)Methanol-F-silicene (e) Methylacetate-F-silicene and (f)Toluene-F-silicene systems.The shallow donor state in each case is marked by arrow.}  \label{Fig:6}
\end{center}
\end{figure*}

The adsorption of VOCs on F-silicene sheet produces remarkable changes in the DOS (Figure \ref{Fig:5}). A common feature found in all six cases is that the projected DOS of the molecules display features of an isolated molecule. The molecular DOS now consists of sharp discrete peaks. This is an artefact of the weak reactivity with the nanosheet due to the protective layer formed by the -F functional group as discussed in the previous sub-section. However, in this case we find significant changes in the DOS of the systems. As mentioned earlier, functionalisation by -F opens up a semiconducting gap. Adsorption of different VOC molecule produces discrete molecular states (contributed by the valence states of the constituents aligned closer to the surface of the nano-material) very close to the Fermi levels of the systems. The position of this isolated state creates a noticeable difference between electronic structures of different VOC-F-silicene complexes. This feature can be quite useful from the perspective of differences in charge transfers and sensitivity when F-silicene is used as a sensor material. The band structures of the six compounds presented in Figure \ref{Fig:6} demonstrate this more clearly. In all cases, an isolated flat band appears below the Fermi level. This band represents the highest occupied molecular orbital (HOMO) of a given complex. This acts as a shallow donor state \cite{zhang2015first}, reduces the band gap of the systems substantially, rendering the systems like a n-type semiconductor. The new band gap $E_{n}$, calculated as the energy difference between the conduction band minimum and the HOMO, is 0.055, 0.053, 0.281, 0.195, 0.346 and 0.137 eV for Acetone,Dimethylsulfide,Ethanol,Methanol,Methylacetate and Toluene adsobed F-silicene, respectively. Such considerable reductions in the band gap imply lower activation energies required for the electrons in the shallow donor states for charge transport, an advantage from the perspective of transport applications.   
\begin{figure}[t]
\begin{center}
\includegraphics[width=0.48\textwidth]{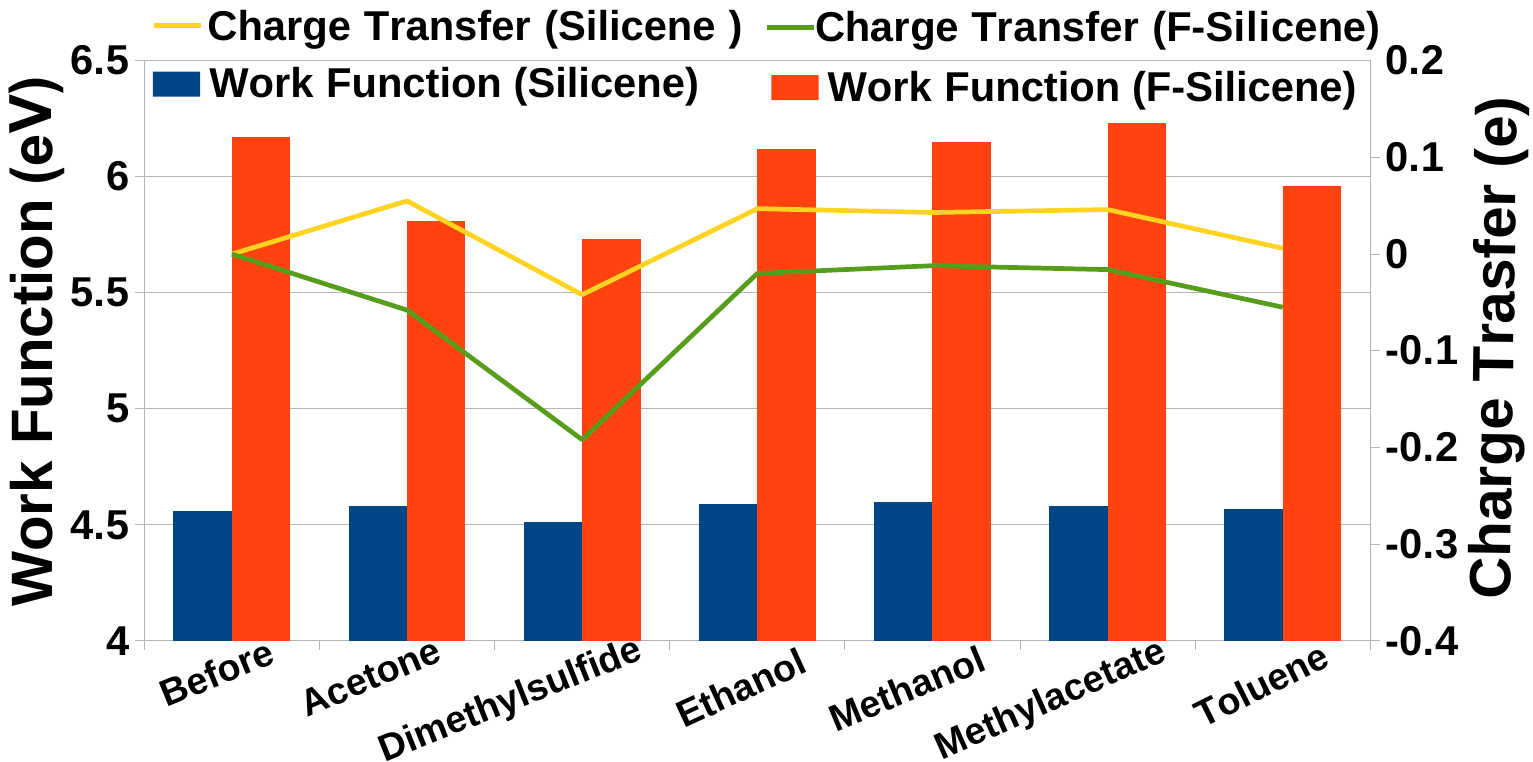}
\caption{Charge Transfer and Work functions before and after adsorption of different VOCs on silicene and F-silicene.}  \label{Fig:7}
\end{center}
\end{figure}
\subsection{Charge transfer and Work function}
In order to substantiate the inferences drawn from analysis of the electronic structures of the VOC-nanosheet complexes, in Figure \ref{Fig:7}, we present the results of charge transfer $Q$ and work function $\phi$  for VOCs adsorbed on both silicene and F-silicene nanosheets. The charges on each component of the complexes are obtained from Bader charge analysis \cite{bader}. The work function $\phi$ is computed from $\phi=E_{\text{vac}}-E_{F}$; $E_{\text{vac}}$ and $E_{F}$ are the vacuum energy and Fermi energy in each case, respectively. $Q <(>)0$ indicates charge transfer from the molecule(nanosheet) to the nanosheet (molecule). 
\begin{figure*}[t]
\begin{center}
\includegraphics[width=0.9\textwidth]{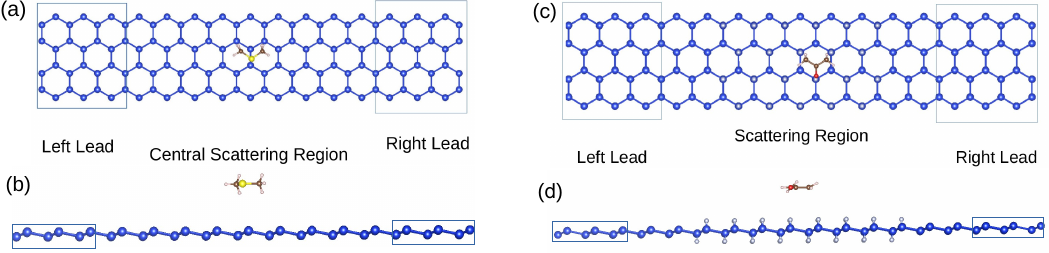}
\caption{(a) Top and (b) side views of the schematic structural model of a gas sensor based on silicene with two semi-infinite electrodes . The areas marked by square boxes represent semi-infinite electrodes.The central part of the figure where the molecule is physisorped (Dimethylsulfide is shown here) is the scattering region. (c) and (d) represent the same but for F-silicene gas sensor. The physisorbed molecule shown in the schematics is Acetone.}  \label{Fig:8}
\end{center}
\end{figure*}
 Analysing the trends in $Q$, we find that the charge transfers, in general, are small, irrespective of the nanosheet. This is consistent with the fact that the adsorption in the present context is physiorption only. The other noticeable features are the following: (a) $Q<0$ when the nano-sheet is F-silicene. This corroborates the donor levels obtained in the band structures. (b)With silicene as the nanosheet, $Q>0$ for all cases except adsorption of Dimethylsulfide. Dimethylsulfide has 2 lone pairs coming from S. As a result it acts as a donor instead of acceptor. (c) $Q$ is greater in magnitude when the nano-sheet is F-silicene. This is consistent with the fact that the molecular electronic states lie deeper in energies (Figure \ref{Fig:4}) when molecules are adsorbed in pristine silicene. The greater charge transfer for F-silicene is also an artefact of the presence of highly electronegative -F on the surface of the 2D material. (d) With silicene as the adsorbent, there is hardly any charge transfer between Toluene and silicene ($Q=0.006e$) while it is maximum in case of Acetone adsorbate ($Q=0.055e$). For the other four molecules, $Q$ varies between 0.042 and 0.047. This corroborates the implications stated out by analysis of the DOS. (e) For both adsorbents, dimethylsulfide enforces the largest charge transfer. (f) In case of F-silicene nanosheet, charge transfer amounts are almost identical when Acetone and Toluene are adsorbed ($Q=-0.058e, -0.055e$ for Acetone and Toluene, respectively). Qualitatively the same is seen when the other three VOCs are adsorped. However, $Q$ is less in their cases: -0.012e, -0.020e and -0.016e for Methanol, Ethanol and Methylacetate, respectively. Once again, such qualitative behaviour validates the conclusions from the electronic structures. It is to be noted that the appreciable changes in the charge transfers upon adsorption of different molecules on an adsorbate are supposed to affect it's resistance that can be experimentally measured and is directly connected to the sensitivity when used as a sensor. 

In an adsorbate-adsorbent complex, if the absorbate is a acceptor (donor), the Fermi energy of the system is lowered( elevated) increasing (decreasing) the work function as a consequence. Work function $\phi$ of pristine silicene (F-silicene) is 4.56 eV (6.17 eV). Higher $\Phi$ of F-silicene is due to strong Si-F bonds as opposed to unpassivated silicene surface. From Figure \ref{Fig:7}, we find that $\Phi$ of the compounds after adsorption of VOCs on pristine silicene increase only slightly (4.57 eV for Toluene adsorped to 4.61 for Methanol adsorped silicene); the only exception is Dimethylsulfide where $\phi$=4.51 eV. This is understandable as the molecule is a donor. Large changes in $\phi$ are observed when the adsorbent is F-silicene. The changes are hardly noticeable (with respect to $\phi$ of F-silicene before adsorption) for Methylacetate, Ethanol and Methanol. Large changes are observed in case of other three molecules: $\Phi$=5.73 eV, 5.82 eV and 5.96 eV for Dimethylsulfide, Acetone and Toluene, respectively. This trend of $\phi$ correlates with $Q$; larger the $Q$, lower is the work function. The results, therefore, indicate possible selectivity of these three molecules when F-silicene is used as a sensing material. On the other hand, the results imply that silicene, when used as a sensor for these VOCs, may not be able to discriminate between them as has been inferred from the electronic structure.
\begin{figure}[t]
\begin{center}
\includegraphics[width=0.485\textwidth]{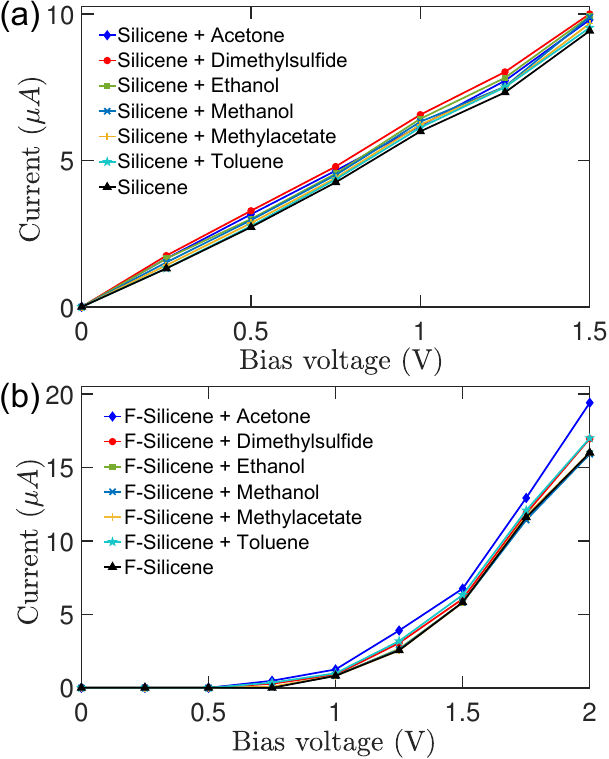}
\caption{I-V characteristic of (a)silicene and (b)F-silicene sensors with and without adsorption of the six VOCs separately.}  \label{Fig:9}
\end{center}
\end{figure}
\subsection{Sensing performances}
Understanding of sensing capabilities of silicene based gas sensors is done by modelling the nano-device. We have considered a two-probe model where the left and right electrodes act as the source and drain of electrons. They are basically extensions of the central scattering region where adsorption of a VOC molecule takes place. As a result, their chemical potentials are same ($\mu_{L}=\mu_{R}$). In Figure \ref{Fig:8} (a) and (b), we show the device setup with  pristine and F-silicene, respectively. The central region is smoothly joined to the electrodes that are periodic in the transport direction ($z$ in this case). To screen out the perturbations in the sensing regions, the electrode regions are extended into the central scattering region. The $I-V$ characteristics with and without gas adsorptions by silicene and F-silicene are shown in Figure \ref{Fig:9} (a) and (b), respectively. Application of a bias voltage $V_{b}$ to the device elevates the Fermi level of the left electrode with respect to the right electrode. When $V_{b}$ supersedes the threshold voltage at which conduction band minimum of the right electrode matches with the valence band maximum of the left electrode, current starts to flow. As seen in Figure \ref{Fig:9}, this threshold is 0V(0.5 V) for pristine silicene (F-silicene). The values are consistent with the electronic band gaps of the bare materials. $I-V$ characteristics curves for VOC adsorped silicene are linear while those for F-silicene are parabolic. In pristine silicene device, the curves are barely separable from one another. The current for VOC adsorbed silicene is always higher than that of bare silicene. The current reaches a maximum of 10 $\mu$A at 1.5 V when Dimethylsulfide is adsorbed. The lowest current is observed in case of Toluene adsorption.     
  
\begin{figure*}[t]
\begin{center}
\includegraphics[width=0.9\textwidth]{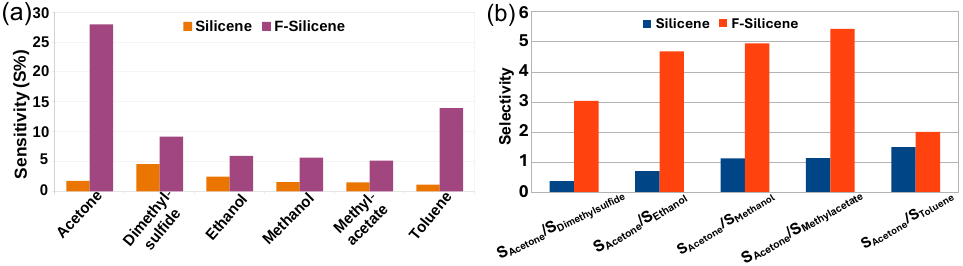}
\caption{(a) Sensitivity of silicene and F-silicene towards the six VOCs considered in this work (b) Selectivity of silicene and F-silicene for detection of Acetone.}  \label{Fig:10}
\end{center}
\end{figure*}

For F-silicene device, the current curves corresponding to each VOC-complex starts to get separated from each other around a bias voltage of 0.75V. For each VOC-F-silicene complex, maximum of current is higher than the one corresponding to pristine silicene device. This suggests better sensing capability of F-silicene as predicted in earlier sub-sections. In here, highest current is observed for Acetone-F-silicene system with a maximum of 19.4 $\mu$A at 2V. Toluene and Dimethylsulfide are the two recording next highest currents. The $I-V$ curves for the other three VOC complexes are barely separable from one another and the bare F-silicene. This behaviour is consistent with the qualitative trends in charge transfer and work functions. 

Direct assessment of sensing capability and one to one comparison with experimental results is done by calculating sensitivity in terms of differential conductance. The sensitivity $S$(\%) is given by 
\begin{align}
S= \frac {|g-g_{0}|}{g_{0}} \times 100
\end{align}
where $g$($g_0$) is the zero-bias conductance of the sensor material when VOC molecule is adsorbed(without the molecular adsorption). With a finite bias $V_{b}$, the conductance can be expressed as $g \left(V_b \right)=\frac{2e^2}{h}T\left(\mu=E_{F}-eV_{b} \right)$; $T$ is the transmission coefficient, $E_F$ the Fermi energy, $h$ the Planck's constant and $e$ the electronic charge. Applied bias voltage $V_{b}$ can effect changes in the transmission by tuning the chemical potential $\mu$. Conductance is a better measure than current because (i) the result is directly comparable to experiments where they measure differential resistance and (ii) it allows a better distinction of the change of signal as the conductance through the sensor changes by a quantum unit of $2e^2/h$ in its interaction with the VOC. 

In Figure \ref{Fig:10}(a), we compare the sensitivities of silicene and F-silicene sensors with regard to the six food VOC molecules. The results presented are for $V_{b}=1.23$ V. For silicene gas sensor, the maximum sensitivity of only about 5 \% is obtained for Dimethylsulfide. The lowest sensitivity is obtained for Toluene ($\sim 1$ \%). The sensitivities for Acetone, Methanol and Methylacetate are almost same. The sensitivity of Ethanol is slightly higher. The trend is consistent with the behaviour of charge transfer for VOC-silicene complexes. A remarkable improvement is obtained in case of F-silicene sensor. The sensitivity for Acetone increases nearly 16 times reaching near 30\%. The sensitivities of Toluene and Dimethylsulfide too improves significantly reaching near 14 \% and 10 \% respectively. For the rest three VOCs too, sensitivities improve to near 5\%. The sensitivities with regard to Acetone and Toluene, obtained in our calculations, are either better or comparable with respect to results obtained with other 2D sensors. For example, sensitivity for Acetone and Toluene with BC$_{6}$N sensor are 6.1 \% and 14.7 \% while those with defective-BC$_{6}$N are 1.7 \% and 8.8 \% only \cite{vocbc6n}. For Ethanol and Methanol, sensitivity of F-silicene sensor is somewhat poor as compared to sensitivity obtained with other 2D materials. With BC$_{6}$N, the sensitivity for Ethanol and Methanol are 14.7 \% and 9.5 \% respectively. It is 61 \% and 8.8 \% with defective BC$_{6}$N, 21 \% for both with MoSe$_{2}$ \cite{mose2} and 9 \% for Ethanol with Black Phosphorous \cite{ou2019superior}  

In Figure \ref{Fig:10}(b), we show comparative abilities of silicene and F-silicene sensors in discriminating the molecules. Sensitivity of Acetone has been considered the reference and the ratio of it's sensitivity to that of another molecule is considered quantitative measure of selectivity of that molecule. The selectivity using silicene sensor is poor in comparison with F-silicene. Infact, silicene sensor cannot separate Methanol and Methylacetate at all. On the contrary, F-silicene can clearly distinguish between Acetone, Dimethylsulfide, Toluene and Methylacetate. The selectivities of Ethanol and Methanol are very close, 4.6 and 4.9 respectively. Therefore, we can conclude that F-silicene can separate out at least four VOCs emitted by food products. The performance is significantly better than r-GO sensors \cite{meatvoc} which could not discriminate between any of the six VOCs. 

\subsection{Recovery Time}
The re-usability of a sensor is dependent on how fast the desorption of a gas molecule takes place. This is quantified by sensor recovery time given as 
\begin{align}
\tau = \nu_0^{-1}{\rm exp }( -E_{\text{ad}} / k_B T ),
\end{align}
$\nu_0$ is the operating frequency, $T$ the temperature, and $k_B$ the
Boltzmann constant. In order for a gas sensor to be reusable $\tau \lesssim 10^{5}$s \cite{recovery}. In Table \ref{TABLE2}, we present calculated values of $\tau$ for both silicene and F-silicene sensor. 
\begin{table}
\begin{tabular}{c|cc}
\hline
Molecule & silicene & F-silicene \\ \hline
Acetone & $1.15 \times 10^{-7}$ & $3.89 \times 10^{-7}$ \\ 
Dimethylsulfide & $2.29 \times 10^{-5}$ & $7.57 \times 10^{-7}$ \\
Ethanol & $4.38 \times 10^{-7}$ & $1.58 \times 10^{-8}$ \\
Methanol & $3.27 \times 10^{-8}$ & $1.13 \times 10^{-9}$ \\
Methylacetate & $5.95 \times 10^{-6}$ & $1.2 \times 10^{-6}$ \\
Toluene & $0.18$ & $2.85 \times 10^{-4}$\\ \hline
\end{tabular}
\caption{Recovery time $\tau$ (in s)of silicene and F-silicene gas sensors to sense food VOCs. The calculations are done at room temperature ($T=300$K) and under visible light ($\nu_{0}=10^{12}$ s$^{-1}$) }
\label{TABLE2}
\end{table}
The results suggest that both silicene and F-silicene can work as reusable sensors for sensing the VOCs from food materials. Calculated recovery times for silicene based sensors turn out to be smaller than those for pristine and defected- BC$_{6}$N. $\tau$ of pristine(defected) BC$_{6}$N sensors in case of  Acetone,Ethanol,Methanol and Toluene  desorption are $2.6 \times 10^{-6} (740)$ s, $2.3 \times 10^{-6} (4.9)$ s, $2.1 \times 10^{-7} (4.2)$ s and $2 \times 10^{3} (6.8 \times 10^{3})$ s, respectively, orders of magnitude higher than silicene based sensors. Thus the silicene and F-silicene nanosheets will function more efficiently as sensors as compared to many other 2D materials.
\section{Conclusions}
Usage of nano-sensors to detect quality of food products is a relatively new addition in the field of nano-science and nanotechnology. Using a combination of DFT based first-principles method and non-equilibrium Green's function technique, we have investigated in detail the sensing abilities of two-dimensional silicene and -F functionalised silicene nanosheets for standard VOCs emitted by canonical food products like meat, fruits and vegetables. We have systematically analysed the structural parameters, the energetics, the electronic structures and charge transfers when silicene based 2D materials are used to adsorb the VOC molecules. Based upon these, we have concluded that  F-silicene to be a better candidate as sensors. Our insights from such analysis of microscopic physics are then put to test by modelling sensor devices and making a direct comparison with the experimental quantities. We find that F-silicene indeed has a much better sensitivity than pristine silicene as far as VOCs from food products are concerned. The sensitivities for select molecules obtained by using F-silicene as the sensing material compare well with the available results using other 2D materials as sensors. Moreover, silicene-F is able to clearly discriminate between at least four molecules, Acetone, Dimethylsulfide, Methylacetate and Toluene. The selectivity is comparable for the other two. A comparison with the experimental findings with r-GO as sensor material \cite{meatexpt} shows that F-silicene is better in discriminating between the VOCs from food materials while r-GO fails to do so. Finally, we establish the re-usability of silicene based sensors to detect these VOCs by computing the desorption time under excitation by visible light. We find that  physiorption of the adsorbates actually help in faster desorption of molecules. This in turn should enhance the sensor's re-usability \cite{usability}. This work carries immense relevance on two larger counts: first, to our knowledge this is the first work using first-principles based computations to investigate sensing performances of 2D materials in the field of food technology, and second, this work opens up possibilities to model 2D materials based sensors to detect and separate more complex VOC molecules emitted from specific food products. For example, VOCs like 3-Carene, Longifolene and D-limonene are found experimentally in Papaya fruits \cite{papayavoc}. However, no e-nose sensor using these VOCs as non-invasive bio-markers to determine the ripening stage of Papaya is yet available. Our study can be extended for such cases in future.  

\section*{Acknowledgement}
	The authors gratefully acknowledge the Department of Science and Technology, India, for the computational facilities under Grant No. SR/FST/P-II/020/2009 and IIT Guwahati for the PARAM ISHAN and PARAM KAMRUPA supercomputing facility where all computations are performed. SG would like to thank Prof. Ramgopal Uppaluri and Ms Paushali Mukherjee, IIT Guwahati, for useful discussions.

\end{document}